\documentclass[aps,amsfonts,pra,singlecolumn,showpacs]{revtex4}
\usepackage{epsfig,amsmath,amssymb,bm,epsf,graphics,psfrag}

\def\bra#1{\langle#1\vert}
\def\ket#1{\vert#1\rangle}
\def\ketbra#1{\vert#1\rangle\langle#1\vert}

\def\ipr#1#2{\langle#1\vert#2\rangle}

\def\Longarrow{\protect\@lra}
\def\@lra{\relbar\joinrel\relbar\joinrel\relbar\joinrel%
          \relbar\joinrel\rightarrow}

\def\tr{{\rm Tr}}

\begin{document}
\title{Entanglement under the renormalization-group transformations on quantum states and in quantum phase transitions}
 \author{Tzu-Chieh Wei}
 \affiliation{
    Institute for Quantum Computing and
    Department of Physics and Astronomy,
    University of Waterloo,
    200 University Avenue West,
    Ontario, Canada N2L 3G1}
\date{July 15, 2009}
\begin{abstract}
We consider quantum states under the renormalization-group (RG)
transformations introduced by Verstraete et al. [Phys. Rev.
Lett. {\bf 94}, 140601 (2005)] and propose a quantification of entanglement  under such RG (via the geometric
measure of entanglement).
We examine the resulting entanglement under RG for the ground states of ``matrix-product-state'' (MPS)
Hamiltonians constructed by Wolf et al. [Phys. Rev. Lett. {\bf
97}, 110403 (2006)] that possess quantum phase transitions.
 We find that near critical points, the
ground-state entanglement
 under RG exhibits singular behavior.
The singular
behavior under finite steps of RG obeys a scaling hypothesis and reveals the correlation length exponent. However, under the infinite
steps of RG transformation, the singular behavior is rendered different and is universal only when
there is an underlying conformal-field-theory description of the critical point.
\end{abstract}
\pacs{03.65.Ud, 03.67.Mn, 64.70.Tg, 11.10.Gh}
 \maketitle
 \section{Introduction}
 Since Wilson~\cite{Wilson}, renormalization-group has been an important tools for theoretical physics,
 ranging from high-energy physics to condensed matter~\cite{Goldenfeld}. It is related to the
 coarse-graining procedure of the physical system, and from that the transformation of system
 parameters in the Hamiltonian is derived. Corresponding terms in the
 Hamiltonian be can determined to be relevant or irrelevant under the scale
 transformation.
 The renormalization-group transformation on quantum states was recently
 introduced by Verstraete et al.~\cite{VerstraeteCiracLatorreRicoWolf05} using the representation of Matrix Product
 States (MPS)~\cite{FannesNachtergaeleWerner92} (for a review of MPS, see e.g. Ref.~\cite{Perez-GarciaVerstraeteWolfCirac07}).
 Many important quantum states in the quantum information theory emerge naturally from the fixed points of
 this coarse-graining transformation on states~\cite{VerstraeteCiracLatorreRicoWolf05}.

One important property associated with quantum states is their entanglement content. There have been tremendous advancement
on the understanding of entanglement (both bipartite and multipartite) for the past few decades~\cite{Horodecki4}.
 The notion of entanglement has also
been applied to many-body systems~\cite{AmicoFazioOsterlohVedral08} and especially
systems that possess quantum phase transitions~\cite{Sachdev},
mainly via bipartite measures (e.g., between two spins or between one subsystem and the remaining) such as concurrence~\cite{Osterloh} and entanglement entropy~\cite{VidalLatorreRicoKitaev03}.
Important insight has thus been obtained, such as the connection to conformal theory
near criticality~\cite{VidalLatorreRicoKitaev03} and to the bipartite entanglement~\cite{Osterloh,WuSarandyLidar06}.
Furthermore, in higher dimensions than one, the entanglement entropy has been shown to
possess an area law for various systems~\cite{Boson,Stabilizer,wolf} (up to logarithmic
corrections in fermions~\cite{wolf}).
 Regarding the coarse-graining process, one can thus raise the interesting question
 how the entanglement behaves under the RG transformation on states~\cite{TagliacozzoOliveiraIblisdirLatorre07}.
  As we shall see below, one of the measures that appear
 to be suitable to the discussions of entanglement under RG is the so-called geometric measure of entanglement (or
 simply geometric entanglement)~\cite{WeiGoldbart03,Wei05}. This measure of
 entanglement
 defined with respect to partitions into blocks of consecutive sites has recently been
 employed by Or\'us~\cite{Orus08} and by Botero and Reznik~\cite{BoteroReznik}.

In this paper, we provide an interpretation of the block geometric entanglement, namely that it is exactly the entanglement
under the coarse graining of the renormalization group transformation on quantum states~\cite{VerstraeteCiracLatorreRicoWolf05}.
We apply this block-$L$ entanglement (with $L$ being the number of sites in each block) to two
spin models constructed by Wolf et al.~\cite{WolfOrtizVerstraeteCirac06}
that possess quantum phase transitions. We find that near critical points, the
ground-state entanglement
 under RG transformation exhibits singular behavior.
The singular
behavior reveals the correlation length exponent. However, under the infinite
steps of RG transformation, the singular behavior is rendered different and it has no universal form
unless the critical point can be described by a conformal field theory.

Let us begin by discussing matrix product states.
It was shown by Vidal~\cite{Vidal03} that any state can be written in the MPS
form as follows,
 \begin{equation}
 \ket{\psi}=\sum_{p_1=1}^{d_1}\sum_{p_2=1}^{d_2}...\sum_{p_m=1}^{d_m}{\rm
 Tr}(A^{[1]}_{p_1}A^{[2]}_{p_2}\cdots A^{[m]}_{p_m})\ket{p_1,p_2,\cdots,p_m},
 \end{equation}
 as long as the matrices $A^{[k]}_{p_k}$'s have large enough dimensions $D_{k-1}\times D_{k}$.
 For any one-dimensional translationally invariant state, all the $A$'s are identical,
 and the state can be written as
\begin{equation}
\label{eqn:psiMPS}
 \ket{\psi}=\sum_{p_1,p_2,...,p_m=1}^{d}{\rm
 Tr}(A_{p_1}A_{p_2}\cdots A_{p_m})\ket{p_1,p_2,\cdots,p_m},
 \end{equation}
 where $A$'s are $D\times D$ matrices with dimension $D\le d^{m/2}$.
 Let us now define the geometric measure of entanglement, by considering a
multipartite system comprising $m$ parts, each of which can have a distinct
Hilbert space. We compare this general $m$-partite entangled pure state
$\ket{\psi}$ to the set of general product pure states,
\begin{equation}
\ket{\phi}\equiv\mathop{\otimes}_{i=1}^m|\phi^{(i)}\rangle,
\end{equation}
and define
 the maximal overlap of $\ket{\psi}$ with the closest product states as
 follows,
\begin{equation}
\Lambda_1({\psi})=\max_{\phi}|\ipr{\phi}{\psi}|.
\end{equation}
The maximal overlap $\Lambda_1(\psi)$ reveals the entanglement content of the
state $\ket{\psi}$, the larger $\Lambda_1(\psi)$, the lower the entanglement of
$\ket{\psi}$. A quantitative way to define the entanglement content is via
\begin{equation}
E_1(\psi)\equiv -\log\Lambda_1({\psi})^2,
\end{equation}
where the subscript $1$ indicates that the product state is composed of product of states of single sites.
Note that the norm square of the translation invariant matrix product state (with $m$ sites)
can be expressed in terms of the operator $\hat{E}$ defined below in Eq.~(\ref{eqn:E}),
\begin{equation}
\ipr{\psi}{\psi}={\rm Tr}({\hat{E}}^m),
\end{equation}
which, for convenience, may not usually be normalized to be unity, and thus
one has to supply this factor in $\Lambda(\psi)$,
\begin{equation}
\Lambda^2(\psi)=\max_{\ket{\phi}\in {\rm
Prod}}\frac{|\ipr{\phi}{\psi}|^2}{\ipr{\psi}{\psi}}.
\end{equation}

We remark that by appropriate partitioning of $\ket{\phi}$ into various
product forms, a hierarchy of entanglement can be
obtained~\cite{BarnumLinden01,WeiGoldbart03,ShimoniBiham07,BlasoneDellAnnoDeSienaIlluminati07}.
The most relevant kind of partitioning regarding RG is to divide $m$ sites
into blocks of several consecutive neighboring sites, e.g., $L$ consecutive sites in
one-dimension. This leads to what we shall refer to as the block-$L$ entanglement,
\begin{equation}
\label{eqn:blockL}
E_L(\psi)\equiv -\log\Lambda_L({\psi})^2,
\end{equation}
where
\begin{equation}
\Lambda_L({\psi})=\max_{\Phi_L}|\ipr{\Phi_L}{\psi}|,
\end{equation}
with $\ket{\Phi_L}$ being product states of the block form:
\begin{equation}
\ket{\Phi_L}\equiv \ket{\phi^{[1..L]}}\otimes\ket{\phi^{[(L\!+\!1)..2L]}}\otimes\cdots,
\end{equation}
and we have implicitly assumed that the total number of spins $m$ (sometimes denoted by $N$) is an multiple of $L$.
\section{RG on quantum states}
 Verstraete et al.~\cite{VerstraeteCiracLatorreRicoWolf05} considered a quantum coarse-graining procedure by merging
 two neighboring spins to one new block spin, which is described in terms of
 matrices $A$'s:
 \begin{equation}
 \tilde{A}^{(pq)}_{\alpha\gamma}\equiv\sum_{\beta=1}^D [A_{p}]_{\alpha\beta}
 [A_{q}]_{\beta\gamma}.
 \end{equation}
From this a more convenient representation is to choose the new matrix as
\begin{equation}
A_{p}\rightarrow A^{'}_l=\lambda_l V^l,
\end{equation}
where $V^l$ is the right unitary matrix in the singular-value decomposition
\begin{equation}
\tilde{A}^{(pq)}_{\alpha\gamma}=\sum_{l=1}^{\min(d^2,D^2)}U^{\dagger(pq)}_l
\lambda_l V^l_{(\alpha\gamma)}.
\end{equation}
This keeps the dimension of the Hilbert space of the block to be of size
bounded above by $D^2$. They further introduced a more convenient
representation of the RG transformation by defining (which we shall thereafter
refer to as the RG operator or the transfer-matrix operator)
\begin{equation}
\label{eqn:E} \hat{E}\equiv\sum_{p=1}^d A_{p}\otimes
A_{p}^*=\sum_{p=1}^d\sum_{\alpha,\beta,\mu,\nu=1}^D
[A_{p}]_{\alpha\beta}\otimes [A_{p}^*]_{\mu\nu}\ket{\alpha\mu}\bra{\beta\nu},
\end{equation}
which is invariant under any local unitary $A_q\rightarrow \sum_{p}U_p^q A_p$.
Hence, the operator $\hat{E}^2$ is invariant under any unitary within the block,
\begin{equation}
{\hat{E}}^2\equiv \sum_{pq}\tilde{A}^{(pq)}\otimes {\tilde{A}^{(pq)*}}=\sum_l
\lambda_l V^l \otimes \lambda_l {V^l}^*.
\end{equation}
 The RG transformation on the state
can then be described by the mapping
\begin{equation}
\hat{E}\rightarrow \hat{E}'={\hat{E}}^2.
\end{equation}

The above discussions assume that the states being considered are translation invariant.
A straightforward extension of the renormalization-group transformations on generic quantum states
can be made by including the site-dependence of the matrices $A_p^{[i]}$ and $E^{[i]}$.
Then one can merge two spins at sites $2k-1$ and $2k$ and form a new spin at the $k$-th site on the new
lattice:
\begin{equation}
 \tilde{A}^{[k](pq)}_{\alpha\gamma}\equiv\sum_{\beta=1}^D [A_{p}^{[2k-1]}]_{\alpha\beta}
 [A_{q}^{[2k]}]_{\beta\gamma},
 \end{equation}
and
\begin{equation}
{{\hat{E'}}}^{[k]}=\hat{E}^{[2k-1]}\cdot \hat{E}^{[2k]}.
\end{equation}
As above, $\hat{E}'$ is invariant under any unitary on the original two spins. This means that under one-step RG transformation, the state $\ket{\psi}$
transforms to
\begin{equation}
\label{eqn:RG2}
\ket{\psi}\rightarrow \ket{\psi'}=U[12]\otimes U[34]\otimes...\otimes
U[2k-1,2k]\otimes...\ket{\psi},
\end{equation}
and the $(2k\!-\!1)$-th and $2k$-th sites are merged into a single site. The unitaries are now generally
site-dependent. Under
RG of merging two neighboring sites,
\begin{equation}
{\rm RG}_2:\ket{\psi}\rightarrow \ket{\psi'}.
\end{equation}
\section{Entanglement of states under RG}
 In a similar spirit, though not identical, to the work on multiscale entanglement renormalization
 ansatz (MERA) by Vidal~\cite{Vidal07}, we discard the short-range description and hence
the total entanglement should decrease under this coarse graining. As the RG
defined by Verstraete et al.~\cite{VerstraeteCiracLatorreRicoWolf05}
transforms the original state to the state $\ket{\psi'}$ up to local (treating
sites $2k\!-\!1$ and $2k$ as local) unitary transformations [see Eq.~(\ref{eqn:RG2})], the natural
definition of the entanglement after the RG is to determine the one that is
minimum among the local equivalence class. In terms of Vidal's MERA, the untiaries $U[2k-1,2k]$ act
as disentanglers that aim to reduce the entanglement between sites $2k-1$ and $2k$. The merging of
the two sites here under the RG defined by Verstraete et al.~\cite{VerstraeteCiracLatorreRicoWolf05}
is done with the same pairs of sites, in contrast to MERA, where the merging is done with, e.g.,
sites $2k$ and $2k+1$ via isometries~\cite{Vidal07}. In conforming the former picture of RG on states, the entanglement
after one-step of RG should be defined as follows:
\begin{equation}
E_G(\{\psi'\})= \min_U E_G(\psi'),
\end{equation}
where the unitary $U$ is of the form $U[12]\otimes U[34]\otimes\cdots\otimes
U[2k-1,2k]\otimes\cdots$.
The maximal overlap in obtaining the entanglement is
\begin{equation}
\max_{\phi's,U}\Big|\Big(\mathop{\otimes}_{i=1}^N\bra{\phi^{[i]}}\Big)\, U
\ket{\psi}\Big|=\max_{\phi's,U}\Big|\Big(\mathop{\otimes}_{k=1}^{N/2}\bra{\,U^\dagger[2k\!-\!1,2k]\phi^{[2k\!-\!1]}\otimes\phi^{[2k]}}\Big)\,
\ket{\psi}\Big|=\max_{\Phi}|\ipr{\Phi}{\psi}|,
\end{equation}
where $\ket{\Phi}=\ket{\phi^{[12]}}\otimes\ket{\phi^{[34]}}\otimes\cdots\otimes\ket{\phi^{[2k-1,2k]}}\otimes\cdots$ with
$\ket{\phi^{[2k-1,2k]}}$ being any arbitrary state of the two sites $2k-1$ and $2k$.
 This is exactly the block-2
entanglement defined previously in Eq.~(\ref{eqn:blockL}) with $L=2$.
One can continue the merging procedure, and arrive at the successive entanglement under RG being
equal to $E_L(\psi)$ with $L=4, 8, \dots, 2^l$, etc.
It it straightforward
to see that for $l'\ge l$, $E_{2^{l'}}(\psi)\le E_{2^l}(\psi)$, i.e., the total entanglement
under RG cannot increase. As another important ingredient in the RG is the rescaling of the lattice
spacing, we also introduce
entanglement per block~\cite{Orus08,BoteroReznik,Wei05} to reflect this
rescaling of length scale (equivalently, system size) in RG,
\begin{equation}
{\cal E}_L(\psi)\equiv \frac{1}{N/L} E_L(\psi),
\end{equation}
where $N$ is the total number of sites (usually considered in the limit of
large number of blocks, $n\equiv N/L\rightarrow \infty$). Therefore, $E_L$ is the
total entanglement for the system with merging $L$ sites into one, and
correspondingly, ${\cal E}_L$ is the entanglement per block.

A conclusion from the above discussions is that the entanglement per block (of size $2^l$) is equal to the entanglement
 per site of the  $l$-time RG transformed state of $\ket{\psi}$, as the RG transformed state is determined
 up to a $2^l$-local sites in view of the original sites.
That is
\begin{equation}
{\cal E}_L(\psi)={\cal E}_1\Big({\rm RG}_2^{\otimes\log_2 L}(\psi)\Big).
\end{equation}
This gives a physical meaning to the block-$L$ geometric measure of
entanglement.

Orus~\cite{Orus08} has recently shown that the geometric measure of
entanglement defined relative to blocks (of size $L$) of spins can be
evaluated via
\begin{equation}
|d_{\max}|^2=\max_{\vec{r}} | (\vec{r}\otimes\vec{r}^*)^\dagger
\hat{E}^L(\vec{r}\otimes\vec{r}^*)|, \ \ \mbox{with}\, |\vec{r}|=1,
\end{equation}
and
\begin{equation}
{\cal E}=-\lim_{n\rightarrow\infty}\frac{1}{n}\log\frac{|d_{\max}|^{2n}}{{\rm
Tr}(\hat{E}^{nL})}=-\log(|d_{\max}|^2),
\end{equation}
where the important assumption is that the closest product state can be taken to be
a product of identical local states. For the ground states of the transverse-field XY spin chains,
this ansatz has been verified numerically~\cite{Wei05}.
Let us briefly describe the proof. Consider the ansatz product state to be
$\ket{\Phi}=\ket{\phi}^{\otimes m}$ and the translation-invariant state
$\ket{\psi}$ is expressed in the MPS form~(\ref{eqn:psiMPS}). Then the overlap
between the two states is
\begin{eqnarray}
&&\ipr{\Phi}{\psi}=\sum_{p_1,p_2,...,p_m=1}^{d}{\rm
 Tr}(A_{p_1}A_{p_2}\cdots
 A_{p_m})\ipr{\phi}{p_1}\ipr{\phi}{p_2}\cdots\ipr{\phi}{p_m}\\
 && ={\rm Tr}(B^m),
\end{eqnarray}
where the matrix $B$ is defined as
\begin{equation}
B\equiv \sum_p A_p \ipr{\phi}{p}.
\end{equation}
Suppose the largest eigenvalue of matrix $B$ is not degenerate, then
\begin{equation}
\lim_{m\rightarrow\infty}{\rm Tr}(B^m)=\lambda_1^m.
\end{equation}
Suppose the corresponding eigenvector is $\vec{r}$, then the goal is to find
\begin{eqnarray}
|\lambda_1|&=&\max_{\vec{r}}| \vec{r}^\dagger B \vec{r}|\\
&=&\max_{\vec{r}}\big\vert\sum_{p}\sum_{\alpha,\beta} r_\alpha^*[A_p]_{\alpha\beta}
\ipr{\phi}{p}r_\beta\big\vert.
\end{eqnarray}
The maximal overlap becomes (in the limit $m\rightarrow\infty$ and taking into account
normalization $\ipr{\psi}{\psi}$)
\begin{equation}
\Lambda(\psi)=\max_\phi\max_{\vec{r}}\Big\vert\sum_{p}\sum_{\alpha,\beta}
r_\alpha^*[A_p]_{\alpha\beta} \ipr{\phi}{p}r_\beta\Big\vert^m/|{\rm
Tr}(\hat{E}^{m})|^{1/2}.
\end{equation}
Under the assumption that the two maximizations can be interchanged, we get
\begin{equation}
\Lambda(\psi)=\max_{\vec{r}}\max_\phi\Big\vert\sum_{p}\sum_{\alpha,\beta}
r_\alpha^*[A_p]_{\alpha\beta} \ipr{\phi}{p}r_\beta\Big\vert^m/|{\rm
Tr}(\hat{E}^{m})|^{1/2}.
\end{equation}
The maximization over $\phi$ can be achieved when (using the Cauchy-Schwartz
inequality), up to a normalization,
\begin{equation}
\ipr{\phi}{p}\sim \sum_{\alpha,\beta} r_\alpha[A_p]_{\alpha\beta}^*r_\beta^*.
\end{equation}
This leads to
\begin{eqnarray}
\Lambda(\psi)^2&=&\max_{\vec{r}}\Big\vert\sum_{p}\sum_{\alpha,\alpha',\beta,\beta'}
r_\alpha^* r_{\alpha'}[A_p]_{\alpha\beta}[A_p]_{\alpha'\beta'}^* r_\beta
r_{\beta'}^*\Big\vert^{m}/|{\rm Tr}(\hat{E}^{m})|\\
&=&\max_{\vec{r}}\Big\vert\sum_{\alpha,\alpha',\beta,\beta'} r_\alpha^*
r_{\alpha'}\hat{E}_{\alpha,\alpha';\beta,\beta'} r_\beta
r_{\beta'}^*\Big\vert^{m}/|{\rm Tr}(\hat{E}^{m})|\\
&=&\max_{\vec{r}}\Big\vert \big(\vec{r}^\dagger\otimes\vec{r}^{\,*\dagger} \hat{E}\,
\vec{r}\otimes\vec{r}^{\,*}\big)\Big\vert^{m}/|{\rm Tr}(\hat{E}^{m})|,
\end{eqnarray}
where $\hat{E}$ is defined in Eq.~(\ref{eqn:E}) as the basis operator for state
renormalization. Therefore, the entanglement per site  becomes
\begin{equation}
\label{eqn:E1}
{\cal E}_1(\psi)=\lim_{m\rightarrow\infty}-\frac{1}{m}\log\Lambda(\psi)^2=
\lim_{m\rightarrow\infty}-\frac{1}{m}\log\left[\max_{\vec{r}}\Big\vert
\big(\vec{r}^\dagger\otimes\vec{r}^{\,*\dagger} \hat{E}\,
\vec{r}\otimes\vec{r}^{\,*}\big)\Big\vert^{m}/|{\rm Tr}(\hat{E}^{m})|\right].
\end{equation}
In the case of entanglement per block of size $L$, we can repeat the previous derivation by replacing
$\hat{E}$ by $\hat{E}^L$ and we arrive at the expression
\begin{equation}
\label{eqn:EL}
{\cal E}_L(\psi)=
\lim_{n\rightarrow\infty}-\frac{1}{n}\log\left[\max_{\vec{r}}\Big\vert
\big(\vec{r}^\dagger\otimes\vec{r}^{\,*\dagger} \hat{E}^L\,
\vec{r}\otimes\vec{r}^{\,*}\big)\Big\vert^{n}/|{\rm Tr}(\hat{E}^{nL})|\right].
\end{equation}
Therefore, the entanglement at large block size $L\rightarrow\infty$ depends
on the fixed-point property of the operator $\hat{E}$.
\section{Fixed points}

 Verstraete et
al.~\cite{VerstraeteCiracLatorreRicoWolf05} have also defined the fixed point under the RG transformation on states via
\begin{equation}
 \hat{E}^\infty\equiv \lim_{l\rightarrow\infty}\hat{E}^l.
\end{equation}
We can use their results on the classification of the fixed point and investigate the
behavior of entanglement for generic states. They concluded that in the
generic case, the largest eigenvalue of $\hat{E}$ is nondegenerate and both its left
and right eigenvectors have a maximal Schmidt rank,
\begin{eqnarray}
\hat{E}^\infty=\ket{\Phi_R}\bra{\Phi_L},
\end{eqnarray}
and one can always choose
\begin{eqnarray}
\ket{\Phi_R}=\sum_{i=1}^D\ket{ii}, \ \
\ket{\Phi_L}=\sum_{i=1}^D\lambda_i\ket{ii},
\end{eqnarray}
where $\lambda_i>0$. Then
\begin{eqnarray}
\bra{r}\bra{r^*}\hat{E}^\infty\ket{r}\ket{r^*}=\sum_{i}\lambda_i|r_i|^2,
\end{eqnarray}
and its maximum is $\lambda_1$, the largest of $\{\lambda_i\}$. Furthermore,
$\tr\big[(\hat{E}^\infty)^n\big]=\Big(\sum_i\lambda_i\Big)^n$. Therefore, the
entanglement per block is
\begin{eqnarray}
{\cal
E}_\infty=-\lim_{n\rightarrow\infty}\frac{1}{n}\log\frac{\lambda_1^n}{\Big(\sum_{i=1}^D\lambda_i\Big)^n}=\log{\Big(\sum_{i=1}^D\lambda_i/\lambda_1\Big)}
=-\log \tilde\lambda_1\le
\log D,
\end{eqnarray}
where we have defined the normalized Schmidt coefficients $\tilde\lambda_i\equiv \lambda_i/\sum_k \lambda_k$.
This means that if a many-body state can be represented by a MPS with
dimension $D$, then the largest entanglement per block at the fixed point is
bounded above by $\log D$. We remark that according to Verstraete et
al.~\cite{VerstraeteCiracLatorreRicoWolf05}, the entropy of a block of spins
is exactly twice the entropy of entanglement of $\ket{\Phi_L}$, i.e.,
$S=-2\sum_i \tilde\lambda_i\log \tilde\lambda_i\le 2 \log D$. In short, we have the relation
${\cal E}_\infty \le S/2\le \log D$.

\section{Examples}
In this section, we shall warm up by several example states.

\noindent {\it Example 1\/}. AKLT state~\cite{AKLT}.\\
\begin{equation}
\{A_p\}=\{\sigma_z, \sqrt{2}\sigma_{+},-\sqrt{2}\sigma_{-}\}.
\end{equation}
Orus has performed a detailed analysis for the AKLT
state~\cite{Orus08b}.
The operator $\hat{E}$ is calculated to be (the matrix elements conveniently
expressed in the ``ket'' and ``bra'' notations,
\begin{eqnarray}
\hat{E}=3\frac{1}{\sqrt{2}}(\ket{00}+\ket{11})\frac{1}{\sqrt{2}}(\bra{00}+\bra{11})-1\cdot\frac{1}{\sqrt{2}}(\ket{00}-\ket{11})\frac{1}{\sqrt{2}}(\bra{00}-\bra{11})
-1\cdot\ket{01}\bra{01}-1\cdot\ket{10}\bra{10}.
\end{eqnarray}
Let $\ket{r}\equiv r_0\ket{0}+r_1\ket{1}$ and $\ket{r*}\equiv
r_0^*\ket{0}+r_1^*\ket{1}$, we have
\begin{eqnarray}
\bra{r}\bra{r^*}\hat{E}^L\ket{r}\ket{r^*}=\frac{3^L}{2}+\frac{(-1)^L}{2}.
\end{eqnarray}
Moreover, for a total of $n$ blocks,
\begin{eqnarray}
{\rm Tr}(\hat{E}^{nL})=3^{nL}+3(-1)^{nL}.
\end{eqnarray}
So
\begin{eqnarray}
{\cal E}_L=-\lim_{n\rightarrow\infty}
\log\frac{\big[\frac{3^L}{2}+\frac{(-1)^L}{2}\big]^n}{3^{nL}+3(-1)^{nL}}=\log2-\log\left[1+\big(-\frac{1}{3}\big)^L\right].
\end{eqnarray}

\noindent {\it Example 2\/}. GHZ state.\\
\begin{eqnarray}
&&A_{\pm}=\openone\pm\sigma_z\\
&&\hat{E}=2\ket{00}\bra{00}+2\ket{11}\bra{11}.
\end{eqnarray}
Then,
\begin{eqnarray}
\bra{r}\bra{r^*}\hat{E}^L\ket{r}\ket{r^*}=2^L(|r_0|^4+|r_1|^4)\le 2^L,
\end{eqnarray}
when either $|r_0|=1$ or $|r_1|=1$. Furthermore,
\begin{eqnarray}
\tr(\hat{E}^{nL})=2\cdot 2^{nL}.
\end{eqnarray}
Therefore,
\begin{eqnarray}
E_L= -\log\frac{ (2^L)^n}{2\cdot 2^{nL}}=\log2, \ \ {\cal E}_L=0.
\end{eqnarray}
Note that in this case, there is a degeneracy in eigenvalues of $\hat{E}$, and there
seems to be no problem in carrying out the procedure. Furthermore, no matter how large $L$ is, GHZ
state looks identical (after rescaling the system size), and it possesses a total entanglement $\log 2$,
and hence vanishing entanglement per block in the limit of large number of blocks.

\noindent {\it Example 3\/}. Cluster state.\\
Consider the linear-cluster state described by
\begin{eqnarray}
A_0=\begin{pmatrix} 0 & 0 \cr 1 & 1
\end{pmatrix}, \ \ A_1=\begin{pmatrix} 1 & -1 \cr 0 & 0
\end{pmatrix}.
\end{eqnarray}
The $\hat{E}$ operator is calculated to be
\begin{eqnarray}
\hat{E}=2\ket{00}\bra{--}+2\ket{11}\bra{++}, \ \ \hat{E}^2=4\ketbra{\Phi^+},
\end{eqnarray}
where $\ket{\Phi^+}=(\ket{00}+\ket{11})/\sqrt{2}$. For $L>1$,
\begin{eqnarray}
\bra{r}\bra{r^*}\hat{E}^L\ket{r}\ket{r^*}=2^{L-1}.
\end{eqnarray}
Furthermore, we have
\begin{eqnarray}
\tr(\hat{E}^{nL})=2^{nL},
\end{eqnarray}
and thus arrive at
\begin{eqnarray}
E_L=n\log2, \ \ {\cal E}_L=\log2.
\end{eqnarray}
We note that the above results hold for $L={\rm even}$; see Example 4 and below. In fact for $L=1$, it is known that
the cluster state possesses $E_1=\lfloor N/2\rfloor\log2$, where $N$ is total number of spins~\cite{MarkhamMiyakeVirmani07}.
This means that ${\cal E}_1=(1/2)\log 2$, which is half of ${\cal E}_{L={\rm even}}$.

\noindent {\it Example 4\/}. Anti-ferromagnetic GHZ state.\\
\begin{equation}
\ket{\psi}=\frac{1}{2}(\ket{01010101...}+\ket{10101010...}).
\end{equation}
One caution is that the closest separable state for block-1 is either
$\ket{01010101...}$ or $\ket{10101010...}$, neither of them being translation
invariant. But block-2 is: e.g., $\ket{(01)(01)(01)(01)...}$. Therefore, there
is an even-odd difference. Of course, the entanglement is the same as the
ferromagnetic GHZ state. To avoid this even-odd effect, we will mainly
consider the block size $L$ to be even. In the following, we will provide numerical
evidence in justifying the ansatz we use to calculate the entanglement.
We note that for permutation invariant pure states, the ansatz of the product states
being a tensor product of identical single-site states is well justified; see Ref.~\cite{Robert}.

\section{Entanglement in quantum phase transitions with matrix product states}
Wolf et al. have recently used matrix product states to engineer quantum phase
transitions (QPT) with properties differing from standard
paradigm~\cite{WolfOrtizVerstraeteCirac06} (e.g., analytic ground state energy
and finite entanglement entropy for an infinite half-chain), but still with
diverging correlation length and vanishing energy gap. Since the ground state
depends on the system parameter ($g$) continuously, we shall investigate the
ground-state entanglement properties of the corresponding models Wolf et al.
considered and determine whether the ground-state entanglement can be a
telltale of the corresponding critical points.

\subsection{A spin-$1/2$ model}
\begin{figure}
\psfrag{g}{$g$}
\psfrag{E1}{${\cal E}_1$}
\psfrag{E2}{${\cal E}_2$}
\centerline{\psfig{figure=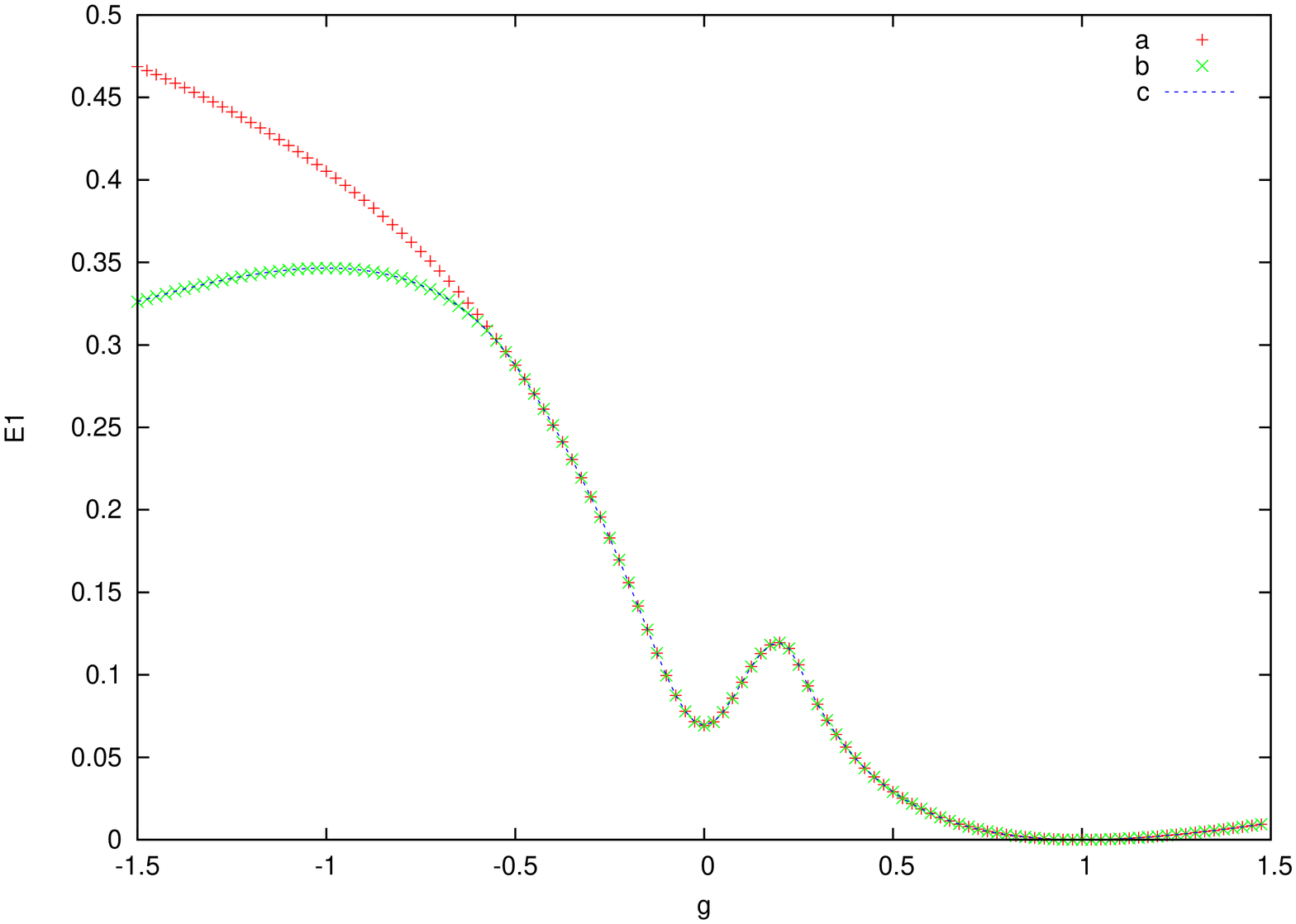,width=7cm,angle=270}}
\vspace{0.5cm}
\centerline{\psfig{figure=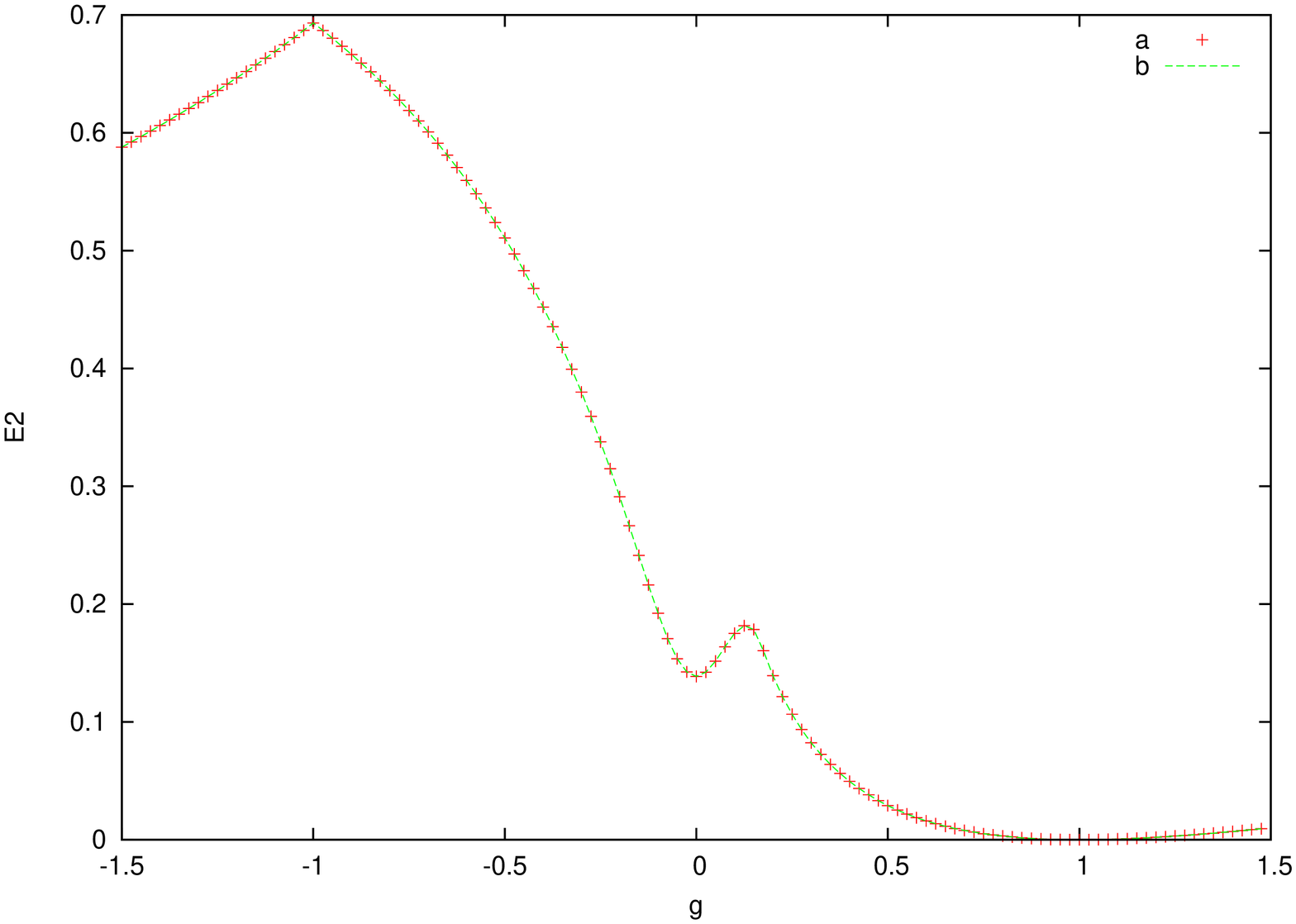,width=7cm,angle=270}}
\caption{Entanglement density ${\cal E}_L(g)$ of the ground state of the
Hamiltonian~(\ref{eqn:H1}) for (a) $N=10$ spins and $L=1$ (top panel); labels ``a'', ``b'', and ``c'' denote the results
from the identical, alternative, and arbitrary ansatz states, respectively; and (b) $L=2$ (bottom panel); here labels ``a'' and ``b'' denote the results
from the identical and arbitrary ansatz states, respectively; }
\label{fig:model1}
\end{figure}
The state that is represented by
\begin{eqnarray}
A_0=\begin{pmatrix} 0 & 0 \cr 1 & 1
\end{pmatrix}, \ \ A_1=\begin{pmatrix} 1 & g \cr 0 & 0
\end{pmatrix},
\end{eqnarray}
is the ground state of the following
Hamiltonian~\cite{WolfOrtizVerstraeteCirac06}
\begin{eqnarray}
\label{eqn:H1} H=\sum_i
2(g^2-1)\sigma_i^z\sigma_{i+1}^z-(1+g)^2\sigma_i^x+(g-1)^2\sigma_i^z\sigma_{i+1}^x\sigma_{i+2}^z.
\end{eqnarray}
At $g=0$, it is a GHZ state, whereas at $g=-1$ it is a cluster state.

\begin{figure}
\psfrag{g}{$g$}
\psfrag{E}{${\cal E}_\infty$}
\centerline{\psfig{figure=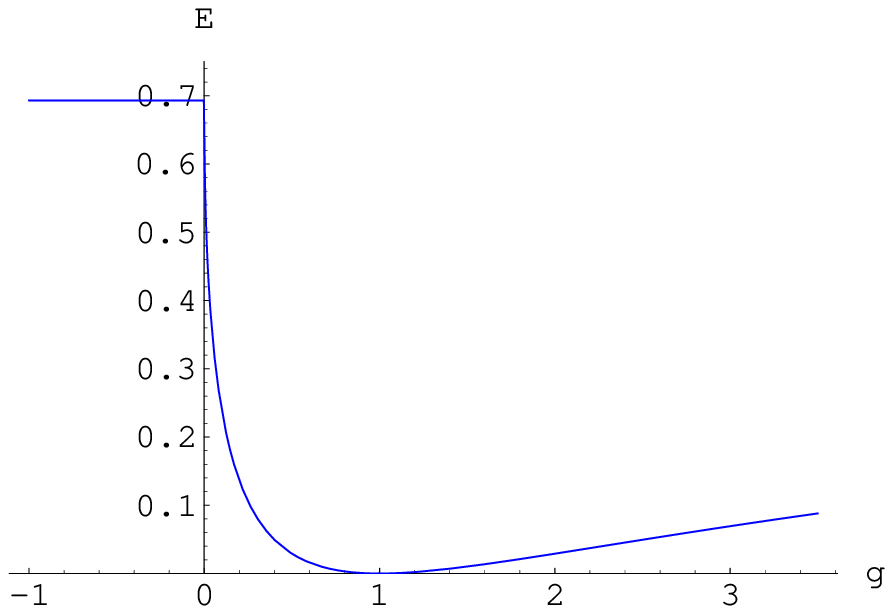,width=8cm,height=5cm,angle=0}} \vspace{0.5cm}
\psfrag{dE}{$d{\cal E}_\infty/dg$}
\centerline{\psfig{figure=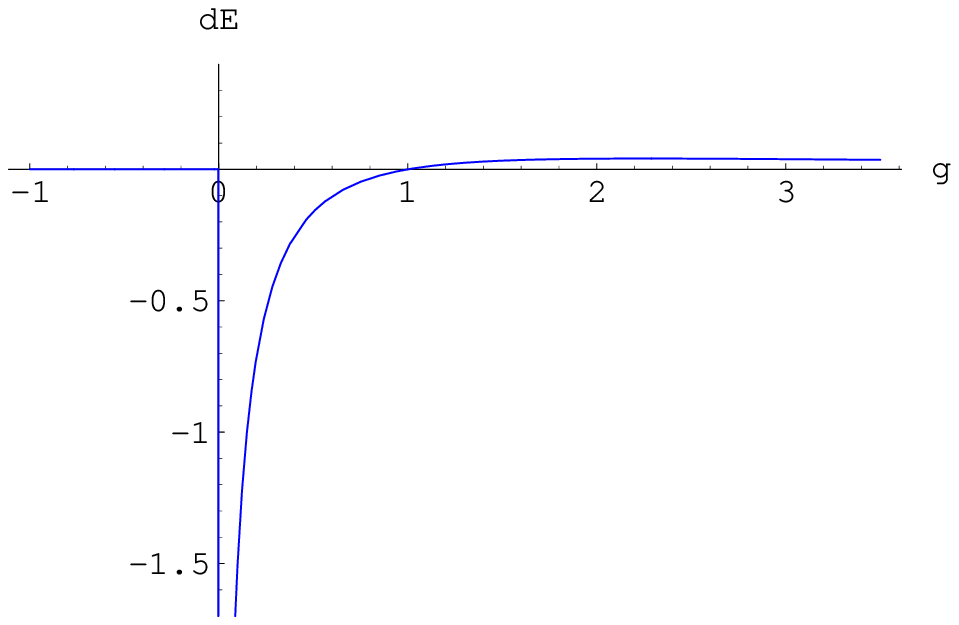,width=8cm,height=5cm,angle=0}}
\caption{Entanglement behavior of the ground state of the
Hamiltonian~(\ref{eqn:H1}). Upper panel: Entanglement ${\cal E}_\infty(g)$.
Lower panel: Entanglement derivative $d{\cal E}_\infty(g)/dg$. }
\label{fig:EntModel1}
\end{figure}

We first check to what extent the ansatz states for deriving the formula~(\ref{eqn:EL}) for ${\cal E}_L$
can be justified. In Fig.~\ref{fig:EntModel1}a, we compare the numerical values of the supposed entanglement
density (${\cal E}_1$) with $N=10$ spins using three different product states (identical,
alternating, arbitrary)
\begin{eqnarray}
\ket{\Phi_a}&=&\ket{\phi^{\otimes N}},\\
\ket{\Phi_b}&=&\ket{\phi_1}\otimes\ket{\phi_2}\otimes\ket{\phi_1}\otimes\ket{\phi_2}\otimes\cdots,\\
\ket{\Phi_c}&=&\otimes_{i=1}^N \ket{\phi_i}.
\end{eqnarray}
We see that the results of using the ansatz $\ket{\Phi_a}$ (identical) to calculate ${\cal E}_1$
are correct only for $g\ge -0.5$. However, the results of using $\ket{\Phi_b}$ (alternating) are always as good as those of
using $\ket{\Phi_c}$ (arbitrary). This suggests that for $L={\rm even}$, an product of
blocks of even number of spins is a good ansatz. This is indeed the case for $L=2$. In Fig.~\ref{fig:model1}b,
we compare numerical results for ${\cal E}_2$ with the same number $N=10$ of spins between two
different ansatz states (identical and arbitrary)
\begin{eqnarray}
\ket{\Phi_a}&=&\ket{\phi_{12}^{\otimes N/2}},\\
\ket{\Phi_b}&=&\ket{\phi_{12}}\otimes\ket{\phi_{34}}\otimes\cdots\otimes\ket{\phi_{N-1,N}}.
\end{eqnarray}
It is clear from the plot that they give identical results, thus supporting the use
of product of identical single block states (with even block size $L$).

Next, we calculate the analytic expression for the entanglement per block, focusing mainly on
$L={\rm even}$. We begin by noting that the operator $\hat{E}$ is
\begin{equation}
\hat{E}=\begin{pmatrix}1& g & g & g^2\cr 0 & 0&0&0\cr 0&0&0&0\cr 1& 1& 1&
1\end{pmatrix}=(1+g)^2\ket{00}\bra{gg}+2\ket{11}\bra{++},
\end{equation}
where $\ket{g}\equiv (\ket{0}+g\ket{1})/\sqrt{1+g^2}$. We can evaluate the
$n$-th power of $\hat{E}$,
\begin{eqnarray}
\hat{E}^n=U\begin{pmatrix}0&0&0&0\cr 0&0&0&0\cr 0&0&(1-g)^n&0\cr
0&0&0&(1+g)^n\end{pmatrix}U^{-1},
\end{eqnarray}
where
\begin{eqnarray}
U=\begin{pmatrix}
  g& 0& -g& g\cr
  -1 - g& -1& 0& 0\cr
  0& 1& 0& 0\cr
  1& 0& 1& 1
 \end{pmatrix}, \ \
 U^{-1}=\begin{pmatrix}
 0& -\frac{1}{1 + g}& -\frac{1}{1 + g}& 0\cr
 0& 0& 1& 0\cr
 -\frac{1}{2 g}& 0& 0& \frac{1}{2}\cr
 \frac{1}{2 g}& \frac{1}{1 + g}& \frac{1}{1 + g}& \frac{1}{2}
 \end{pmatrix}.
\end{eqnarray}
We can then calculate ${\cal E}_L(g)$. In this case, we shall see that even
the behavior of the entanglement ${\cal E}_\infty(g)$ will exhibit singularity
across the critical point $g=0$. We calculate the entanglement ${\cal
E}_\infty(g)$ to be
\begin{eqnarray}
{\cal E}_\infty(g)=\left\{\begin{array}{l}
 \log2+\log(1+g)-2\log{(1+\sqrt{g})}, \ \ \mbox{for} \ g>0,\cr 0,  \ \
\mbox{for} \ g=0,\cr \log2, \ \ \mbox{for} \ g<0.
\end{array}\right.
\end{eqnarray}
The derivative of ${\cal E}_\infty(g)$ is discontinuous across $g=0$ and
exhibits divergence as $g\rightarrow0^+$,
\begin{eqnarray}
\frac{d {\cal E}_\infty(g)}{dg}=-g^{-1/2}, \ \ g\rightarrow 0^+.
\end{eqnarray}
Therefore, the critical point is reflected by the property that the
fixed-point entanglement ${\cal E}_\infty$ has divergence behavior in its
derivative with respect to $g$.

\begin{figure}
\psfrag{g}{$g$}
\psfrag{E}{${\cal E}_L$}
\centerline{\psfig{figure=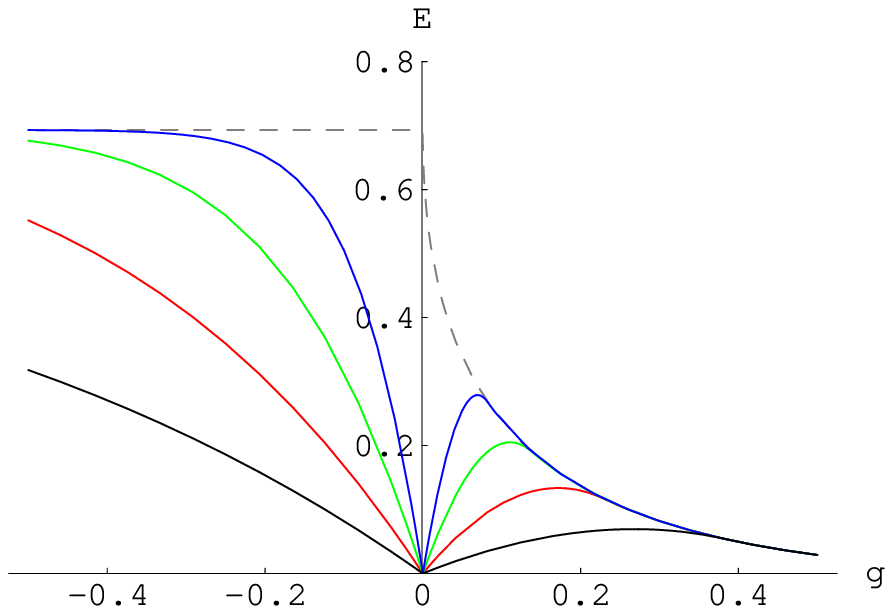,width=8cm,height=5cm,angle=0}}
\vspace{0.5cm}
\centerline{\psfig{figure=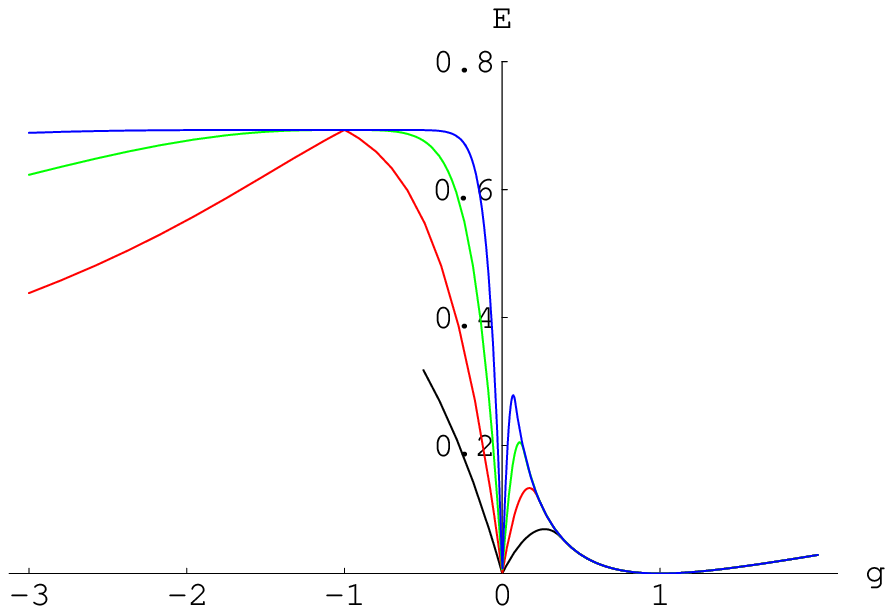,width=8cm,height=5cm,angle=0}}
\caption{Entanglement behavior ${\cal E}_L(g)$ of the ground state of the
Hamiltonian~(\ref{eqn:H1}) for $L=1,2,4,8$ from bottom to top. Note that in the top panel
for $L=1$ only the range $g\ge -0.5$ is shown.
The dashed curve represents the results of ${\cal E}_\infty(g)$.}
\label{fig:EntL1}
\end{figure}

We can also study the entanglement ${\cal E}_L$ for finite $L$ and learn how
the entanglement varies under state-RG transformation. The entanglement for
$L$ finite turns out to exhibit more feature (see Fig.~\ref{fig:EntL1}),
\begin{eqnarray}
{\cal E}_L(g)=\left\{\begin{array}{l} \log2+\log(1+g)-2\log{(1+\sqrt{g})}, \ \
\mbox{for} \ g>0, (1\! + \!g)^L|1\! -\! g|^{-L} \sqrt{g} > (1 \!+\! g) \cr
\log2-\log\Big[1+(1-g)^L(1+g)^{-L}+{g}(1+g)^{L-2}|1-g|^{-L}\Big], \ \
\mbox{for} \ g>0, (1 \!+\! g)^L\big|1\! -\! g\big|^{-L} \sqrt{g} \le (1 \!+\! g), \cr
\log2-\log\Big[1+\frac{1}{2}\Big(1+\sqrt{1+\frac{|g|}{(1-|g|)^2}}\Big){\Big\vert1-|g|\Big\vert^L}{(1+|g|)}^{-L}\Big],
\ \ \mbox{for} \ g<0.
\end{array}\right.
\end{eqnarray}
The singularity at the critical point $g=0$ is obvious; see
Fig.~\ref{fig:EntL1}. In particular, we find that
\begin{equation}
\frac{d {\cal E}_L(g)}{dg}\Big\vert_{0^+}=L-\frac{1}{2}, \ \ \frac{d {\cal
E}_L(g)}{dg}\Big\vert_{0^-}=-({L}-\frac{1}{8}).
\end{equation}
As $g$ decreases from a large value, the entanglement decreases to
zero at $g=1$, which is paramagnetic state with all spins pointing in the $x$
direction. It then rises to a local maximum as $g$ further decreases, and
afterwards decreases to zero at the critical point. This intermediate region
becomes smaller as $L$ increases and is washed out at the fixed point. The cusp for $L=2$ at $g=-1$ reflects a
highly entangled state, which turns out to be the one-dimensional cluster
state. The entanglement slowly decreases as $g$ become more negative.

\begin{figure}
\psfrag{g}{$g$}
\psfrag{C}{$C$}
\centerline{\psfig{figure=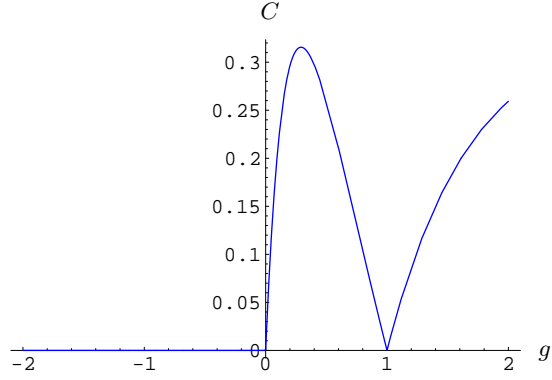,width=8cm,height=5cm,angle=0}}
\caption{Nearest-neighbor Concurrence behavior ${C}(g)$ of the ground state of the
Hamiltonian~(\ref{eqn:H1}). The use of concurrence to infer critical points may incorrectly
identify $g=1$ as a critical point.}
\label{fig:C}
\end{figure}

As a comparison, we show the behavior of the nearest-neighbor concurrence~\cite{Wootters98} in Fig.~\ref{fig:C}.
There are singularities at $g=0$ and $g=1$. The use of concurrence to infer critical points may incorrectly
identify $g=1$ as a critical point.

\subsection{A spin-$1$ model}
\begin{figure}
\psfrag{g}{$g$}
\psfrag{E1}{${\cal E}_1$}
\centerline{\psfig{figure=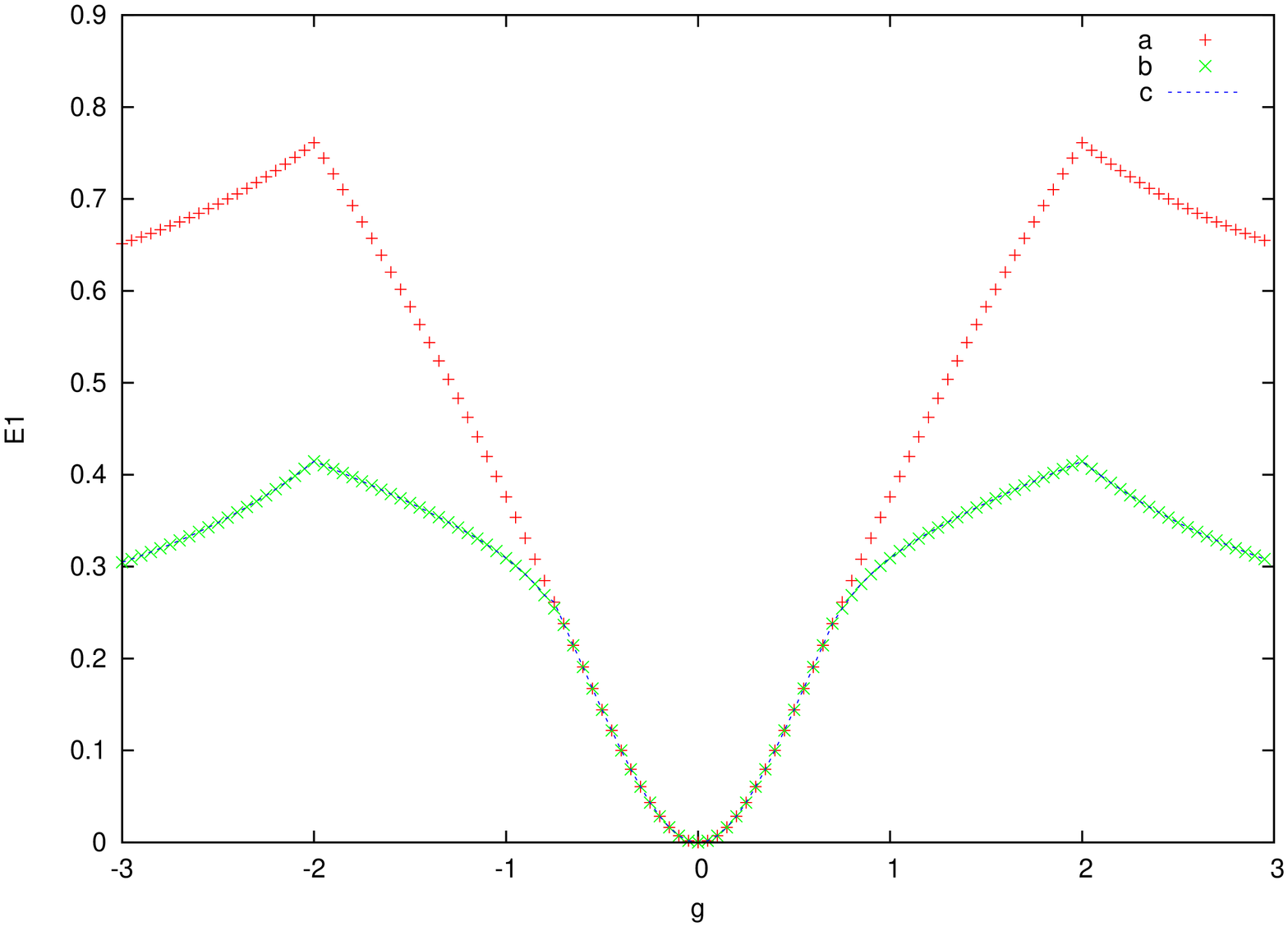,width=7cm,angle=270}}
\vspace{0.5cm}
\centerline{\psfig{figure=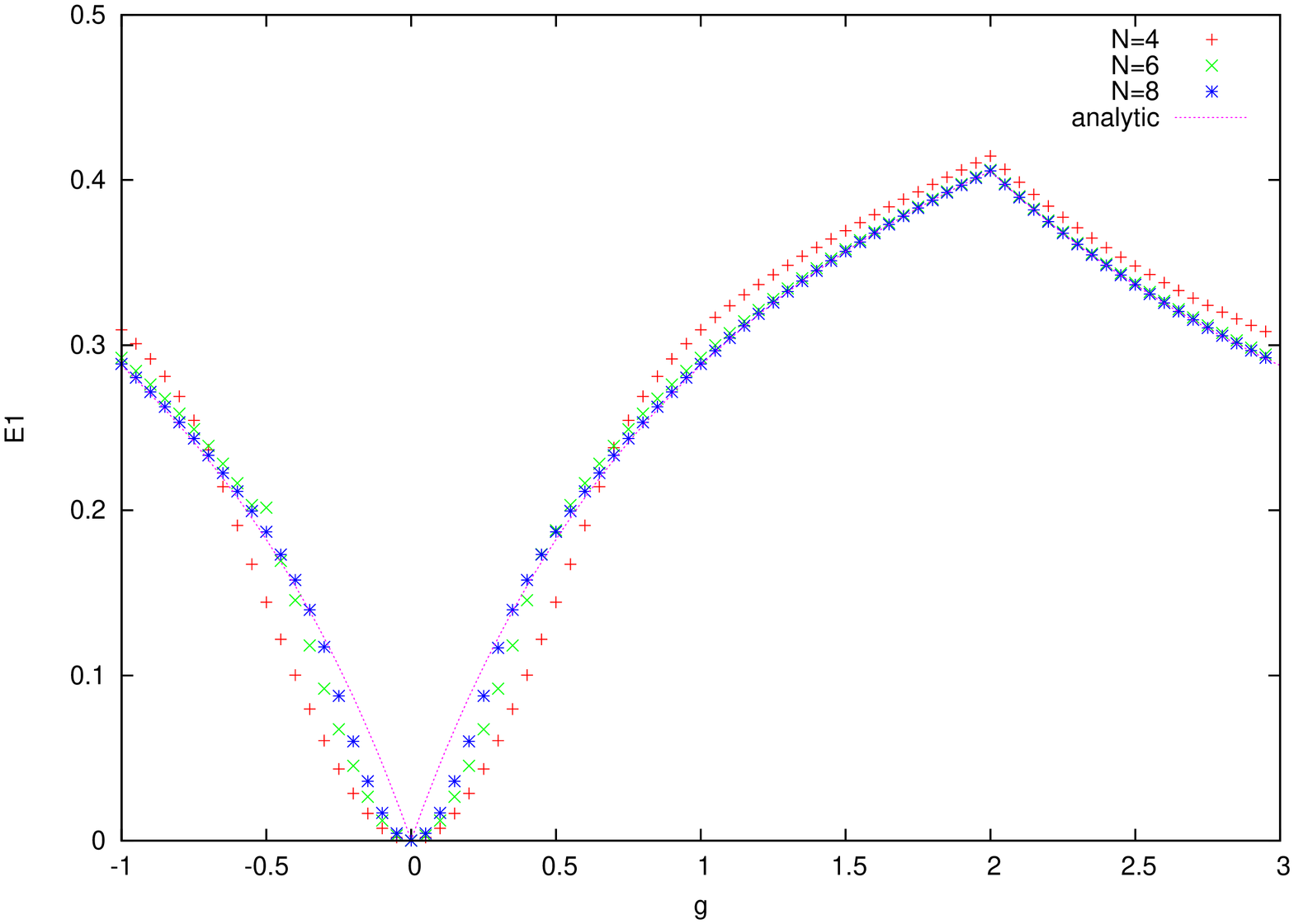,width=7cm,angle=270}}
\caption{Entanglement density ${\cal E}_{L=1}(g)$ of the ground state of the
Hamiltonian~(\ref{eqn:H2}) (a) for $N=4$ (top panel); labels ``a'', ``b'', and ``c'' denote the results
from the identical, alternative, and arbitrary ansatz states, respectively; (b) (bottom panel)
for $N=4,6,8$ and the analytic
expression from Eq.~(\ref{eqn:E_L2}). }
\label{fig:model2}
\end{figure}
This example is a spin-one scenario. The state represented by
\begin{eqnarray}
A_0=-\sigma_z, \ A_1=\sigma^+, \ A_{-1}=g\sigma^-,
\end{eqnarray}
 is the ground state of the following spin-1 Hamiltonian
\begin{eqnarray}
\label{eqn:H2}
H_2&=&\sum_i(2+g^2)\vec{S}_i\vec{S}_{i+1}+2(\vec{S}_i\vec{S}_{i+1})^2+2(4-g^2)(S_i^z)^2\nonumber\\
&& -(g+2)^2(S_i^zS_{i+1}^z)^2+g(g+2)\{S_i^zS_{i+1}^z,\vec{S}_i\vec{S}_{i+1}\}.
\end{eqnarray}
For $g=\pm2$, the GS is the AKLT state. For $g\rightarrow\pm\infty$, the GS is
the N\'eel GHZ state. The critical point is at $g=g_c=0$, where there is a
diverging correlation length.
\begin{figure}
\psfrag{g}{$g$}
\psfrag{E}{${\cal E}_L$}
\centerline{\psfig{figure=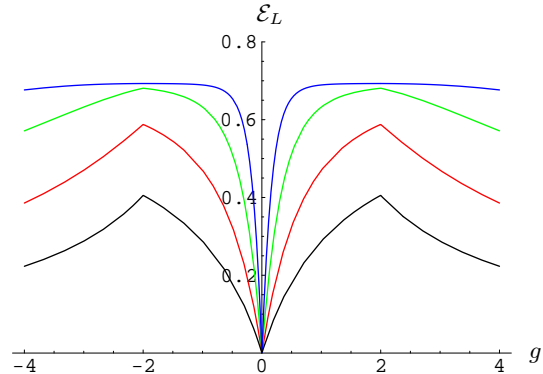,width=8cm,height=5cm,angle=0}}
\caption{Entanglement ${\cal E}_L(g)$ for the ground state of the
Hamiltonian~(\ref{eqn:H2}) for block sizes $L=1,2,4,8$ from bottom to top. } \label{fig:Ent2L}
\end{figure}

Here, we also check to what extent the ansatz states for deriving the formula~(\ref{eqn:EL}) for ${\cal E}_L$
can be justified. In Fig.~\ref{fig:model2}a, we compare the numerical values of the supposed entanglement
density (${\cal E}_1$) with $N=4$ spins using three different product states (identical,
alternating, arbitrary)
\begin{eqnarray}
\ket{\Phi_a}&=&\ket{\phi^{\otimes N}},\\
\ket{\Phi_b}&=&\ket{\phi_1}\otimes\ket{\phi_2}\otimes\ket{\phi_1}\otimes\ket{\phi_2}\otimes\cdots,\\
\ket{\Phi_c}&=&\otimes_{i=1}^N \ket{\phi_i}.
\end{eqnarray}
We see that the results of using the ansatz $\ket{\Phi_a}$ (identical) to calculate ${\cal E}_1$
are correct only for $|g|$ being small. For larger $|g|$ the ansatz $\ket{\Phi_a}$ gives incorrect
results. However, the results of using $\ket{\Phi_b}$ (alternating) are always as good as those of
using $\ket{\Phi_c}$ (arbitrary). Similar to previous model, this suggests that for $L={\rm even}$, an product of
blocks of even number of spins is a good ansatz. We note that for finite $N$ there is a region
near $g=0$ that the entanglement behaves quadratically with $g$. This region, however, shrinks
as $N$ becomes larger, as illustrated in Fig.~\ref{fig:model2}b. In the thermodynamic limit ($N\rightarrow\infty$) this region is expected
to disappear and the behavior of entanglement near $g$ becomes linear (see below). Furthermore,
even though the formula for ${\cal E}_L$ is expected to work for $L={\rm even}$, the expression
obtained by taking $L=1$ appears to be close to the entanglement density obtained for finite $N$,
except the small quadratic region (due to finite system size).

Now, we derive the analytic expression for ground-state entanglement per block.
First, we can construct the renormalization operator $\hat{E}$,
\begin{eqnarray}
\hat{E}=\begin{pmatrix}1& 0 & 0 & g^2\cr 0 & -1&0&0\cr 0&0&-1&0\cr 1& 0& 0&
1\end{pmatrix},
\end{eqnarray}
which can be transformed into the diagonal form
\begin{eqnarray}
\hat{E}=U\begin{pmatrix}-1&0&0&0\cr 0&-1&0&0\cr 0&0&1-g&0\cr
0&0&0&1+g\end{pmatrix}U^{-1},
\end{eqnarray}
where
\begin{eqnarray}
U=\begin{pmatrix} 0& 0& -g& g\cr 0& 1& 0& 0\cr
 1& 0& 0& 0\cr
  0& 0& 1& 1
 \end{pmatrix}, \ \
 U^{-1}=\begin{pmatrix}
 0& 0&1& 0\cr
 0& 1& 0& 0\cr
 -\frac{1}{2 g}& 0& 0& \frac{1}{2}\cr
 \frac{1}{2 g}&0& 0& \frac{1}{2}
 \end{pmatrix}.
\end{eqnarray}
 From this, we can get the ${\cal E}_\infty(g)$,
\begin{eqnarray}
{\cal E}_\infty(g)=\left\{\begin{array}{l} \log2, \ \ \mbox{for} \ g\ne0,
|g|<\infty,\cr 0, \ \ \mbox{for} \ g=0, |g|\rightarrow\infty.
\end{array}\right.
\end{eqnarray}
The critical point in the limit of the fixed-point $L\rightarrow\infty$ is
reflected only by the discontinuity of the entanglement. The more interesting
analysis comes to the case $L$ is finite. For simplicity, take $L$ to be even.
For $|g|\gg 1$, ${\cal E}_L(g)=0$. We find that ${\cal E}_L(-g)={\cal
E}_L(g)$, as can be seen from the symmetry of $\hat{E}$ under $g\rightarrow -g$. For
finite values of $g$, we have
\begin{eqnarray}
\label{eqn:E_L2}
{\cal E}_L(g)=\left\{\begin{array}{l}
\log2-\log\big[1+(1+g)^{-L}(g-1)^{L}\big], \ \ \mbox{for} \ 2<g\cr
 \log2-\log\big[1+(1+g)^{-L}\big], \ \ \mbox{for} \ 0\le g<2. \ \
\end{array}\right.
\end{eqnarray}
We find that there is a discontinuity in the derivative across the critical
point,
\begin{equation}
\frac{d {\cal E}_L(g)}{dg}\Big\vert_{0^+}-\frac{d {\cal
E}_L(g)}{dg}\Big\vert_{0^-}=-{L}.
\end{equation}
 The critical point is revealed by this singularity (i.e., the discontinuity).
In addition to the critical point, the AKLT state at $g=\pm2$ is signalled by
the cusp in the entanglement for $L=2$ and $4$. The weaker singular behavior
at $g=\pm2$ suggests that it is not a critical point.

\section{Entanglement near criticality}
Near quantum critical points,  the correlation length generally scales as
\begin{equation}
\xi(g)\sim |g-g_c|^{-\nu}.
\end{equation}
For any density functions, such as free energy density and the entanglement
density, when there is no logarithmic singularity we expect the scaling behaves as (i.e., the scaling hypothesis, see, e.g. Ref.~\cite{Goldenfeld})
\begin{equation}
{\cal E}_L(g)\sim\big(\xi/L\big)^{-d},
\end{equation}
where $d$ is the dimension of the system. We then expect that (unless ${d
{\cal E}(g)}/{dg}$ behaves as $\sim\log|g-g_c|$) in general
\begin{equation}
\frac{d {\cal E}_L(g)}{dg}\Big\vert_{g_c^\pm}\sim \pm L^d|g-g_c|^{d\nu-1}.
\end{equation}
In the case of one dimension $d=1$ we consider here, the
discontinuity of the entanglement derivative that we have found for both
models (with $L$ finite) implies that we have $\nu=1$, which is indeed the
case. Furthermore, the discontinuity is proportional to $L$, which is expected as
it is now the smallest length scale.

In the case of transverse-field XX model, near the critical field $g_c=1$ the
entanglement density ${\cal E}_1$ has been shown to behave as~\cite{Wei05}
\begin{equation}
{\cal E}_1^{XX}\sim \sqrt{1-g}, \ \ \mbox{for} \, g\rightarrow 1^-,
\end{equation}
and the entanglement derivative behaves as
\begin{equation}
\frac{d{\cal E}_1^{\rm XX}}{dg}\sim -\frac{1}{\sqrt{1-g}}, \ \ \mbox{for} \, g\rightarrow 1^{-}.
\end{equation}
This is consistent with $d=1$ and $\nu=1/2$ for the XX model.

These results need to be compared to the case of one-dimensional
transverse-Ising spin chain, where the divergence of the entanglement
derivative is logarithmic,
\begin{equation}
\frac{d {\cal E}_{L=1}^{\rm Ising}(g)}{dg}\sim- \frac{1}{2\pi}\log|g-1|, \ \ g\rightarrow
1,
\end{equation}
where $g$ is the ratio between the external field and the spin-spin coupling
and $g=1$ is the critical point. In this case finite-size scaling needs to
be performed to determine $\nu=1$~\cite{Wei05}.

We remark that the divergence behavior  in ${\cal E}_L$  can be rendered
different as $L\rightarrow\infty$ (i.e., much larger than the correlation length
$\xi$), so the scaling hypothesis is no longer expected to hold.
The two MPS models discussed above already illustrate this: ${\cal E}_{{\rm finite}\, L}$ and ${\cal E}_\infty$
have different singular behaviors near critical points. For the transverse-Ising model, the results
of Or\'us~\cite{Orus08}
showed that near $g=1$, the entanglement ${\cal E}_L$ with block size $L\gg \xi$ behaves as
\begin{equation}
{\cal E}_\infty(g)\sim\frac{c}{12}\log\xi\sim\frac{c}{12}\log|g-1|^{-1},
\end{equation}
which gives (where $c$ is the central charge of the conformal field theory)
\begin{equation}
\frac{d {\cal E}_{\infty}(g)}{dg}\sim\mp\frac{c}{12}|g-1|^{-1}, \ \
g\rightarrow 1^\pm.
\end{equation}

In brief, near criticality the
singular behavior of entanglement under finite steps of RG reveals the critical point and
the associated correlation-length critical
exponent $\nu$. However, under the
infinite steps of RG transformation, the singular behavior is rendered
different, which seems to be non-universal for critical points not describable by a conformal
field theory, such as the two MPS models discussed here~\cite{WolfOrtizVerstraeteCirac06}. But it is
universal for critical points describable by a conformal field theory such as the transverse-field Ising model.
\section{Fidelity measure}
\begin{figure}
\psfrag{g}{$g_1$}
\psfrag{f}{$f$}
\psfrag{h}{$g_2$}
\centerline{\psfig{figure=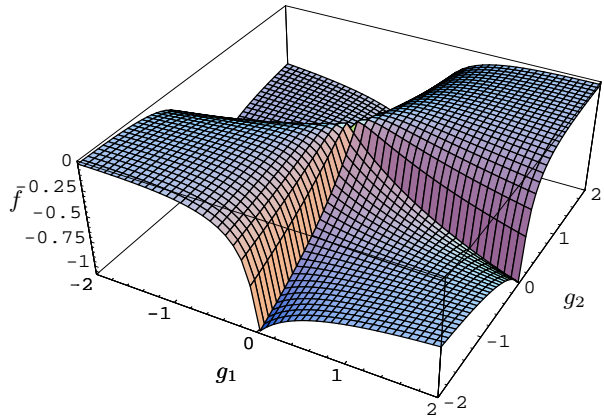,width=8cm,height=6cm,angle=0}}
\caption{Fidelity measure $f(g_1,g_2)$ between ground states of the
same model at two different system parameters $g_1$ and $g_2$.
It turns out this both Hamiltonians~(\ref{eqn:H1}) and~(\ref{eqn:H2}) have
the same plot.  } \label{fig:fidelity}
\end{figure}

As a comparison, we compute the logarithmic fidelity~\cite{fidelity} between ground states at two different system
parameters ($g=g_1$ and $g=g_2$),
\begin{equation}
f(g_1,g_2)\equiv \lim_{m\rightarrow\infty}\frac{1}{m}\log \frac{|\ipr{\psi(g_1)}{\psi(g_2)}|^2}{\ipr{\psi(g_1)}{\psi(g_1)}
\ipr{\psi(g_2)}{\psi(g_2)}},
\end{equation}
where $m$ is the number of sites (or spins). It has been shown that the fidelity measure can
reveal the underlying critical point.
 The inner product between two matrix
product states
\begin{eqnarray}
 \ket{\psi}&=&\sum_{p_1=1}^{d_1}\sum_{p_2=1}^{d_2}...\sum_{p_m=1}^{d_m}{\rm
 Tr}(A_{p_1}A_{p_2}\cdots
 A_{p_m})\ket{p_1,p_2,\cdots,p_m},\\
\ket{\phi}&=&\sum_{p_1=1}^{d_1}\sum_{p_2=1}^{d_2}...\sum_{p_m=1}^{d_m}{\rm
 Tr}(B_{p_1}B_{p_2}\cdots
 B_{p_m})\ket{p_1,p_2,\cdots,p_m},
 \end{eqnarray}
is straightforward to evaluate:
\begin{equation}
\ipr{\phi}{\psi}={\rm Tr}(W^m),
\end{equation}
where $W\equiv \sum_{p_k}A_{p_k}\otimes B^*_{p_k}$. The fidelity measure for both models has been
calculated by Cozzini, Ionicioiu and Zanardi for arbitrary $m$~\cite{Cozzini06}. Here we
compare with the results in the limit of $m\rightarrow\infty$. Interestingly, for both models,
we obtain the same expression
\begin{eqnarray}
f(g_1,g_2)=\left\{\begin{array}{l} \log\frac{(1+\sqrt{g_1g_2})^2}{(1+|g_1|)(1+|g_2|)}, \ \
\mbox{for} \ g_1g_2\ge0, \cr
\log\frac{1+|g_1g_2|}{(1+|g_1|)(1+|g_2|)}, \ \
\mbox{for} \ g_1g_2<0.
\end{array}\right.
\end{eqnarray}
The results are shown in Fig.~\ref{fig:fidelity}. The critical point ($g_1=g_2=0$) is
at the intersecting point of four different regions. As seen from the figure, there is
 singularity  across the lines $(0,g_2)$ and $(g_1,0)$, and $(0,0)$ is the intersection
 of all these four lines.
\section{Concluding remarks}
We have considered the entanglement of states under the renormalization-group
(RG) transformations and apply to the ground state of ``matrix-product-state''
Hamiltonians constructed by Wolf et al. For these models, the entanglement entropy
of a consecutive $L$ spins does not scale logarithmically with $L$. Furthermore, the
use of concurrence for one of the models can lead to a spurious critical point.
Using the geometric entanglement under RG, we have found that near critical
points, the ground-state entanglement exhibits singular behavior. The
singular behavior under finite steps of RG
obeys a scaling hypothesis (similar to free energy) and
reveals the correlation length exponent. However, under
infinite steps of RG transformation, the singular behavior is rendered
different. It is universal only when
there is an underlying conformal-field-theory description of the critical point.
Along the way, we have provided an upper bound for entanglement per
block, which is $\log D$, where $D$ is the dimension of the matrices in MPS.
This also shows the more complex the ground state is, the larger the dimension
of the representative matrices for MPS we need to use.

We conclude by posing the question whether
 there is any significance to the following function
\begin{equation}
\beta(g,L)\equiv \frac{d {\cal E}_L(g)}{d \log L}.
\end{equation}
At least it is a quantity showing how the entanglement changes under the RG
scale transformation; see Figs.~\ref{fig:EntL1} and~\ref{fig:Ent2L}. Under RG
(i.e., as $L$ increases) certain singular but non-critical features get washed
or smoothed away. But the singular behavior near criticality persists for
large $L$. The critical points of the two Hamiltonians~(\ref{eqn:H1})
and~(\ref{eqn:H2}) actually have different fates approaching fixed points. The
former becomes an isolated point in the entanglement, whereas the latter
become an algebraic singularity in the entanglement derivative, albeit both singularities are rendered
different from finite $L$.
\smallskip

\noindent {\it Acknowledgment\/}. The author would like to acknowledge Pochung Chen for useful
remarks regarding matrix product states and Ian Affleck, Gerardo Ortiz, Roman Or\'us and Michael Wolf for useful discussions. This work was supported by IQC, NSERC, and
ORF.

\end{document}